\documentclass[sigconf]{acmart}

\usepackage{pifont}%

\usepackage{amsthm}
\usepackage{amsmath}
\usepackage{amssymb}
\usepackage{tabularx}  
\usepackage{graphicx}
\usepackage{subcaption}
\usepackage{hyperref}
\usepackage{multirow}
\usepackage{enumitem}
\usepackage{hyperref}
\usepackage{algorithm}
\usepackage{algorithmic}
\usepackage{balance}
\usepackage{accents}
\AtBeginDocument{%
  }

\newcommand{\cmark}{\ding{51}}%
\newcommand{\xmark}{\ding{55}}%

\usepackage[HTML]{xcolor}
\usepackage{tikz}

\definecolor{stepblue}{HTML}{5B738B}
\newcommand{\step}[1]{%
Step~%
\tikz[baseline=(char.base)]{%
\node[circle, fill=stepblue, text=white, inner sep=1.00pt] (char) {\textbf{#1}};%
}%
}

\copyrightyear{2026}
\acmYear{2026}
\setcopyright{cc}
\setcctype{by}
\acmConference[WWW '26] {Proceedings of the ACM Web Conference 2026}{April 13--17, 2026}{Dubai, United Arab Emirates.}
\acmBooktitle{Proceedings of the ACM Web Conference 2026 (WWW '26), April 13--17, 2026, Dubai, United Arab Emirates}
\acmISBN{979-8-4007-2307-0/2026/04}
\acmDOI{10.1145/3774904.3792346}

\begin{document}

\title[Diversity-Augmented Negative Sampling for Implicit Collaborative Filtering]
{Diversity-Augmented Negative Sampling for \\Implicit Collaborative Filtering}

\author{Yueqing Xuan}
\affiliation{
  \institution{RMIT University}
  \city{Melbourne}
  \state{Victoria}
  \country{Australia}
}
\email{yueqing.xuan@rmit.edu.au}
\orcid{0000-0002-9365-8949}

\author{Kacper Sokol}
\affiliation{
  \institution{Universit\`{a} della Svizzera italiana}
  \city{Lugano}
  \country{Switzerland}
}
\email{kacper.sokol@usi.ch}
\orcid{0000-0002-9869-5896}

\author{Mark Sanderson}
\affiliation{
  \institution{RMIT University}
  \city{Melbourne}
  \state{Victoria}
  \country{Australia}
}
\email{mark.sanderson@rmit.edu.au}
\orcid{0000-0003-0487-9609}

\author{Jeffrey Chan}
\affiliation{
  \institution{RMIT University}
  \city{Melbourne}
  \state{Victoria}
  \country{Australia}
}
\email{jeffrey.chan@rmit.edu.au}
\orcid{0000-0002-7865-072X}

\renewcommand{\shortauthors}{Yueqing Xuan et al.}



\begin{abstract}
 Recommenders built upon %
 implicit collaborative filtering are typically trained to distinguish between users' positive and negative preferences. %
 When direct observations of the latter are unavailable, negative training data are constructed with sampling techniques. %
But since items often exhibit clustering in the latent space, existing methods tend to oversample negatives from dense regions, resulting in %
homogeneous training data and limited model expressiveness. %
 To address these shortcomings, %
 we propose a novel negative sampler with diversity guarantees. %
 To achieve them, our approach first pairs each positive item of a user %
 with one that they have not yet interacted with; %
 this instance, called \emph{hard negative}, is chosen as the top-scoring item according to the model. %
 Instead of discarding the remaining highly informative items, we store them in a user-specific cache. %
 Next, our diversity-augmented sampler selects a representative subset of negatives from the cache, ensuring its dissimilarity from the corresponding user's hard negatives. %
 Our generator then combines these items with the hard negatives, replacing them to produce more effective (synthetic) negative training data %
 that %
 are informative and diverse. %
Experiments show that our method %
 consistently leads to superior recommendation quality without sacrificing computational efficiency. %
\end{abstract}

\maketitle

\section{Introduction}

Recommender systems learn users' preferences and suggest items of potential interest by leveraging historical user--item interaction data. %
Many leading collaborative filtering recommenders rely on implicit feedback data and adopt a pairwise learning paradigm such as Bayesian Personalised Ranking (BPR)~\cite{rendle2012bpr}. %
These models are typically trained on triplets linking a user with two items, one representing a positive and the other a negative interaction. %
However, in many real-world applications, only positive user feedback -- e.g., clicks, purchases or views -- is available as negative preferences can rarely be observed. %
Consequently, items that users genuinely dislike are often indistinguishable from those they have simply not encountered. %
Negative Samplers (NSs) are then employed to identify the \emph{true negative} items among all the unlabelled ones, %
with
the informativeness of these instances being crucial to the effectiveness of the recommenders trained on them. %

Current negative sampling strategies often follow a two-stage pipeline. First, a small candidate subset of items is uniformly sampled from the full set of unlabelled instances to reduce the downstream computational cost.  %
Then, these instances are ranked according to method-specific criteria, with the top item -- called the \emph{hard negative} -- used for model training. %
For example, Dynamic Negative Sampling (DNS) selects the candidate with the highest predicted score~\cite{zhang2013optimizing}, and 
Disentangled NS (DENS) picks the negative that diverges the most from the positive item along disentangled latent dimensions~\cite{lai2023disentangled}.  %

\begin{figure}[b]
    \centering
    \begin{subfigure}{0.48\linewidth}
        \includegraphics[width=\linewidth]{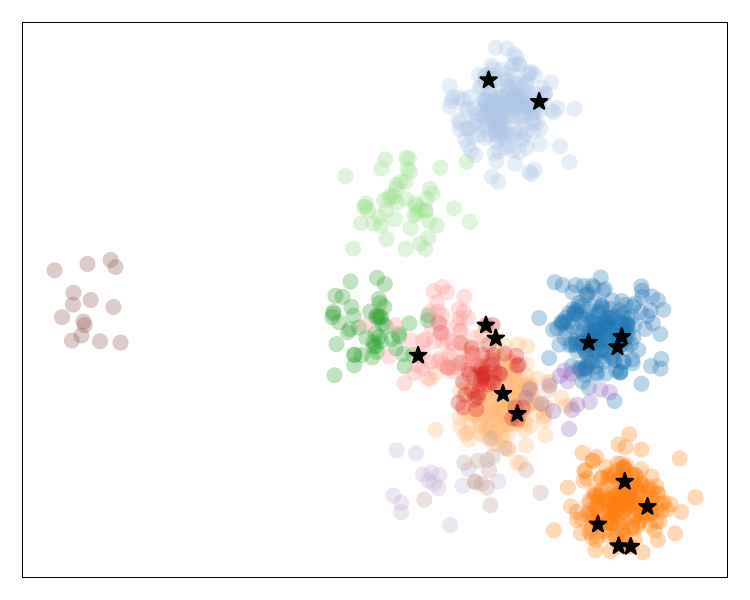}%
        \caption{Uniform sampling.}
        \label{fig:random_sampling}
    \end{subfigure}
    \hfill%
    \begin{subfigure}{0.48\linewidth}
        \includegraphics[width=\linewidth]{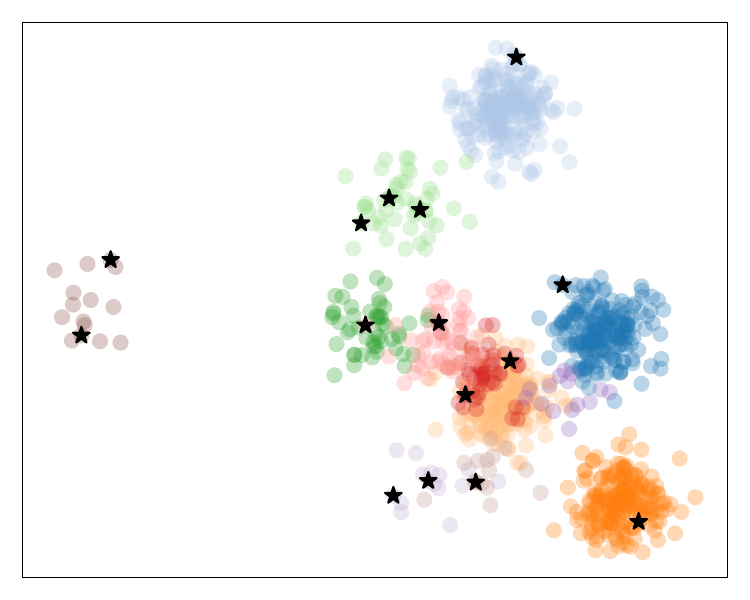}%
        \caption{Sampling based on $k$-DPP.}
        \label{fig:dpp_sampling}
    \end{subfigure}
    \caption{Toy example showing the distribution of $k=15$ negative items ($\star$) drawn through (\subref{fig:random_sampling})~uniform and (\subref{fig:dpp_sampling})~$k$-Determinantal Point Process ($k$-DPP) sampling. %
    Item clusters in the embedding space are marked with different colours.} %
    \label{fig:two_sampling}
\end{figure}

Notably, in the first stage, these methods sample candidate negatives independently for each positive item of a user. 
Due to the inherent cluster structure of item embeddings -- attributed to factors such as item genre or popularity -- uniform sampling often selects neighbouring negatives from the same high-density regions of the embedding space.  %
Figure~\ref{fig:random_sampling} illustrates this issue with a toy example, where the selected candidates are concentrated within a few dense clusters, resulting in homogeneous negative samples. %
When this process is repeated for different positive items, NS tools frequently resample similar negatives from these dense areas, leading to reuse of the same items during model training. %
Meanwhile, sparse regions of the data space -- potentially containing highly informative negatives -- are systematically overlooked. %

As a result, positive items are repeatedly paired with negatives that are similar to one another, leading to limited exploration of the item space during model training. %
This lack of diversity among the negative samples yields uninformative gradient updates when optimising the BPR loss and poor expressiveness of the resulting recommender %
as we show later in %
Section~\ref{sec:motivation}~\cite{duan2022learning,zhang2024harnessing,hong2024diversified}. %
Notably, %
diverse training data have been shown to benefit model learning across various machine learning domains.  %
For instance, \citeauthor{duan2022learning}~\citep{duan2022learning} demonstrated that diverse negative instances improve node classification performance in Graph Neural Networks (GNN), and %
\citet{zhang2024harnessing} illustrated the importance of this property %
for pre-training large language models; similarly, \citet{hong2024diversified} showed that training on diverse data subsets improves model generalisation. %

In recommender systems, however, directly maximising diversity by sampling negatives across the entire item space is impractical. 
Since the item pool is typically overwhelmingly large, na\"ively optimising for this property is computationally infeasible. %
To overcome this limitation, %
we propose a novel %
diversity-augmented negative sampling technique that is tailored to recommender systems and %
generates representative subsets of informative negatives that exhibit beneficial variation without sacrificing the tractability of this process or requiring item labels and repeated item evaluation. %
In particular, we demonstrate that incorporating diversity criteria into the model training process itself -- specifically, the task of selecting negative data samples from the latent space -- can lead to improved model generalisation and more accurate recommendations. %

In contrast to our problem formulation -- which remains largely under-explored -- most recommender research focuses on increasing diversity of the final recommendation output; %
this is often achieved via post-processing techniques such as result re-ranking based on item dissimilarity~\cite{kunaver2017diversity,wu2024result}. %
Related work on instance sampling for model training, on the other hand, is predominantly concerned with data imbalance; %
for example, FairNeg achieves category-aware sampling by adjusting sampling probabilities across item categories to rebalance training data~\cite{chen2023fairly}. %
Notably, %
such methods typically rely on external metadata, e.g., category labels, and do not explicitly optimise for diversity of the sampled negatives in the learnt embedding space. %

Our approach -- called \textbf{Div}erse \textbf{N}egative \textbf{S}ampling (\textbf{DivNS}) -- targets BPR-based recommenders and %
consists of three modules: %
\begin{description}[topsep=0pt]%
\item[cache construction]%
maintains a per-user cache of informative negatives not yet used in model training; %
\item[diversity-augmented sampling]%
selects a diverse subset of negative items from the cache while ensuring that they are dissimilar from current hard negatives, leveraging $k$-Determinantal Point Process ($k$-DPP) to this end; and %
\item[synthetic negatives generation]%
combines the selected diverse negatives with hard negatives to form the final training set. 
\end{description}
This pipeline ensures that negative samples used for training are both informative and representative, i.e., have good coverage of the latent item space. %
Consequently, DivNS enables %
recommenders to learn from a broader region of the item space, thus improving model generalisation and mitigating overfitting to narrow data subspaces.  %
Furthermore, our method is computationally efficient and easy to integrate into existing recommenders without requiring additional item metadata, labelling or repeated scoring. %

In summary, our work delivers three main contributions: %
\begin{enumerate}[leftmargin=*,topsep=0pt]
    \item We identify and theoretically analyse the diversity limitations of existing negative sampling strategies, showing their tendency to oversample from dense regions of the item space.
    \item We propose DivNS -- a novel negative sampling approach that incorporates a caching mechanism, diversity-augmented $k$-DPP sampling and an item mixing strategy to ensure that the resulting instances are both informative and diverse.  %
    \item We conduct an extensive empirical evaluation to demonstrate that incorporating diverse negative samples into training data enables DivNS to consistently outperform strong baselines in terms of recommendation quality. %
\end{enumerate}

\section{Related Work and Preliminaries}

\subsection{Negative Sampling}%

Recommenders are commonly trained on implicit user feedback data, where the observations represent users' positive preference towards items; %
the unobserved data then contain both true negative items and potential positive ones that have not yet been interacted with. %
Negative sampling strategies are used to extract true negative items from the unobserved data. They are generally categorised into \emph{static} and \emph{hard} sampling approaches.

\emph{Static} NS utilises a fixed sampling distribution for each user. %
The simplest approach uniformly samples negative instances from a user's non-interacted items~\cite{rendle2012bpr}; advanced methods introduce more informative sampling distributions. %
For example, MCNS approximates the negative sampling distribution using the positive distribution~\cite{yang2020understanding}, and PNS uses a sampling method based on item popularity to select highly popular instances as negative data~\cite{chen2017sampling}. %

\emph{Hard} NS, on the other hand, targets items that are challenging for the model to distinguish from positives, which typically are instances with high probability of being misclassified as positive interactions for a given user~\cite{zhang2013optimizing,rendle2014improving,wang2017irgan}.  %
These items tend to lie close to decision boundaries, hence using them as negative training data allows more precise delineation of user preferences. %
For instance, DNS dynamically selects negative items with the highest predicted preference scores~\cite{zhang2013optimizing}, and SRNS samples items that have high prediction scores and high score variance as these are considered indicators of true negative instances~\cite{ding2020simplify}. %
MixGCF generates synthetic hard negatives by mixing information from neighbouring items as well as positive instances~\cite{huang2021mixgcf}, and %
DENS disentangles item features based on their relevance, using this contrast observed between positives and candidate negatives to guide sampling~\cite{lai2023disentangled}. %
SCONE injects stochastic positive items to form hard negatives~\cite{lee2025scone}.

Beyond this categorisation, several works propose samplers targeting other recommender setups. %
AdvIR integrates negative sampling into a GAN framework to generate robust negatives~\cite{park2019adversarial}. Auxiliary information such as social relations~\cite{chen2019samwalker} or cross-domain user preferences~\cite{wang2025exploration} are also frequently utilised to guide sampling. For sequential recommendation, GNNO adopts neighbourhood-based hard negative mining~\cite{fan2023neighborhood}.

Despite these advancements, the diversity criterion remains largely overlooked by negative samplers. Existing methods retrieve negatives independently for each positive interaction, often resulting in highly similar negatives being repeatedly matched with the positive items of a given user. %
Our work is the first to explicitly incorporate diversity into negative sampling for recommenders, aiming to %
enhance model generalisation across the item space.

\subsection{Determinantal Point Processes}
Determinantal Point Processes (DPPs) have emerged as a principled probabilistic framework for modelling diversity in subset selection~\cite{kulesza2012determinantal}. A DPP defines a distribution over subsets of items, favouring sets whose elements are dissimilar to each other under a similarity kernel. 
$k$-DPP is a generalisation of DPP for sampling subsets of fixed (rather than variable) size $k$~\cite{kulesza2011k}. %
Although exact $k$-DPP sampling involves eigen-decomposition and is computationally expensive, scalable approximations that use low-rank projections or greedy maximum a posteriori (MAP) inference have been proposed to make this technique fast and practical~\cite{gartrell2017low}. %
Please refer to Appendix~\ref{app:dpp-background} for further details on ($k$-)DPPs. %

In recommenders, %
DPPs are primarily used as post-processing tools to re-rank recommendation results and promote diversity in the final output~\cite{borodin2017max,sha2016framework}. For example, \citet{liu2020diversified} use DPP sampling to improve recommendation diversity in an online setting; %
similarly, \citet{wu2019pd} propose a personalised DPP matrix for generating diverse results.
These methods improve novelty and serendipity of the final recommendation received by the users, but they ignore the lack of diversity in negative data used for model training. %
DPPs, nonetheless, have also the potential to alleviate the latter, %
especially in negative sampling -- a topic that remains under-explored. %

By leveraging $k$-DPPs to construct training negatives that are diverse, we aim to reduce model overfitting to a small set of instances and facilitate more precise decision boundary construction across the item space. %
To the best of our knowledge, our work is the first to introduce diversity-aware negative sampling via DPPs in the recommender training process. %

\subsection{Preliminaries}
Our work focuses on implicit recommendation, where user preferences are inferred from observed interactions rather than explicit ratings.  %
Let $\mathcal{U}$ and $\mathcal{I}$ respectively denote the space of users and items. %
The interaction space is then defined as $\mathcal{Y} =\{(u,i) \; | \; u\in \mathcal{U}, i\in \mathcal{I} \}$, with $(u,i)$ becoming a boolean indicator in a $|\mathcal{U}| \times |\mathcal{I}|$ matrix of whether a user $u$ has interacted with an item $i$.  %
For a given user $u \in \mathcal{U}$, $I_u^+ \subseteq \mathcal{I}$ denotes the set of items they have interacted with, i.e., positive items, where $I_u^+ = \{i \; | \; (u,i) \equiv 1, i\in\mathcal{I}\}$.  Then, $I_u^-=\mathcal{I}\setminus I_u^+$ denotes an item set that the user $u$ has not yet interacted with. 
During recommender training, each item $i \in \mathcal{I}$ is represented by a normalised $d$-dimensional embedding vector $\mathbf{v}_i \in \mathbb{R}^d$ constrained to $\|\mathbf{v}_i\| = 1$. The recommender learns a scoring function $\hat{y}: \mathcal{U} \times \mathcal{I} \rightarrow \mathbb{R}^+$ approximating the preference of $u$ towards $i$. %

In the pairwise learning paradigm, each observed positive interaction $i^+ \in I_u^{+}$ for a user $u \in \mathcal{U}$ is paired with a negative item $i^- \in I_u^{-}$. %
To this end, a negative sampler first draws a candidate set of negative items $\accentset{\star}{I}_u^- = \{i_1^-, i_2^-, \dots, i_n^-\} \subset I_u^{-}$ of size $n$ uniformly from $I_u^{-}$, where $n = |\accentset{\star}{I}_u^-| \ll |I_u^{-}|$.  %
Then, the sampler deploys its own criteria to rank the items in $\accentset{\star}{I}_u^-$ and selects the top one, denoted with $\accentset{\star}{i}_u^-$, as the corresponding hard negative sample; %
for example, DNS formulates this task as $\accentset{\star}{i}_u^- = \arg\max_{i^- \in \accentset{\star}{I}_u^-}{\hat{y}(u,i^-)}$. Then $\accentset{\star}{i}_u^-$ is paired with the underlying positive item $i^+$ to compute pairwise ranking loss %
such as the BPR loss~\cite{rendle2012bpr}, which is defined as: %
\begin{equation*}
    \ell_{\mathit{BPR}} = - \sum_{u \in \mathcal{U}} \sum_{i^+\in I_u^+}
    \ln \sigma\!\left(\hat{y}(u,i^+) -\hat{y}(u,\accentset{\star}{i}_u^-)\right)\text{,}
\end{equation*}
where each $\accentset{\star}{i}_u^-$ is retrieved specifically for every single $i^+$ as explained above and %
$\sigma(\cdot)$ is the sigmoid function. Minimising $\ell_{\mathit{BPR}}$ encourages the predicted score of each positive item $\hat{y}(u,i^+)$ to be higher than that of its paired negative item $\hat{y}(u,\accentset{\star}{i}_u^-)$. %

\section{Homogeneity of Negative Data}\label{sec:motivation}%
Next, we examine the lack of diversity in negative training data produced by current negative samplers. We analyse this problem from both theoretical and empirical angles, and discuss why directly maximising item diversity is infeasible in recommender systems. %

\subsection{Theoretical Analysis}\label{sec:problem-exp}

We define the diversity of a candidate negative sample set $\accentset{\star}{I}_u^-$ using the average pairwise cosine dissimilarity:
\begin{equation}\label{eq:sim-metric}
\begin{aligned}
    \mathit{div}(\accentset{\star}{I}_u^-) & = \frac{1}{n(n-1)} \sum_{i \ne j \in \accentset{\star}{I}_u^-} \left(1 - \cos(\theta_{ij})\right) \\
    & = 1 - \frac{1}{n(n-1)} \sum_{i\ne j \in \accentset{\star}{I}_u^-} \mathbf{v}_i^\top \mathbf{v}_j~\text{,}
\end{aligned}
\end{equation}
where $\theta_{ij}$ is the angle between $\mathbf{v}_i$ and $\mathbf{v}_j$; %
the equality $\cos(\theta_{ij}) = \mathbf{v}_i^\top \mathbf{v}_j$ comes from the aforementioned normalisation condition. %

\begin{figure*}[t]
    \centering
    \includegraphics[clip,trim={7.5pt 0 17.5pt 2.5pt}, width=0.780\linewidth]{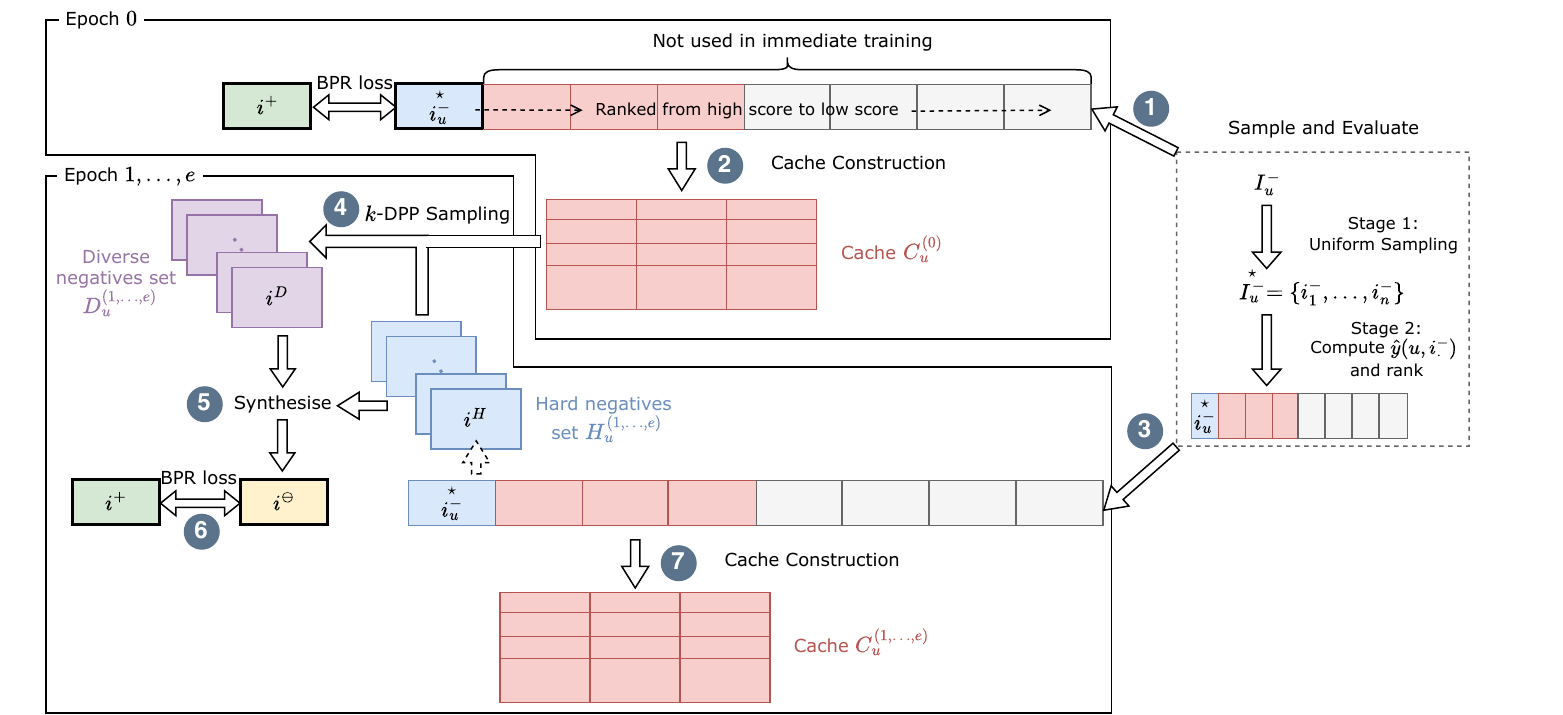}
    \caption{Illustration of DivNS. %
    Epoch 0 is used for initialisation; subsequent epochs -- Epoch 1, \ldots, e -- are identical.}%
    \label{fig:flow-chart}
\end{figure*}

For NS strategies that rely on uniform sampling to generate $\accentset{\star}{I}_u^-$, the latent space embedding of these items $\{\mathbf{v}_i\}_{i \in \accentset{\star}{I}_u^-}$ tends to exhibit a cluster structure in real-world recommender systems due to factors such as item categorisation, genre or popularity. %
Under uniform sampling, the likelihood of selecting multiple items from the same cluster is high. Since items within the same cluster typically have high similarity, we have: %
\begin{equation*}
\mathbb{E}_{\accentset{\star}{I}_u^- \sim \mathit{Uniform}} \left[ \mathbf{v}_i^\top \mathbf{v}_j \right]
> \min \frac{1}{n(n-1)} \sum_{i \ne j \in \accentset{\star}{I}_u^-} \mathbf{v}_i^\top \mathbf{v}_j~\text{.} %
\end{equation*}

It then follows that the expected diversity of randomly sampled negative sets is strictly lower than the maximum achievable one: %
\begin{align*}
    \mathbb{E}_{\accentset{\star}{I}_u^- \sim \mathit{Uniform}}
    [\mathit{div}(\accentset{\star}{I}_u^-)] & = 1 - \mathbb{E}_{\accentset{\star}{I}_u^- \sim \mathit{Uniform}} \left[ \frac{1}{n(n-1)} \sum_{i \ne j \in \accentset{\star}{I}_u^-} \mathbf{v}_i^\top \mathbf{v}_j \right] \\  %
    & < \max \mathit{div}(\accentset{\star}{I}_u^-)~\text{.}
\end{align*}
Therefore, we posit that current NS strategies cannot always guarantee that the candidate negative sets $\accentset{\star}{I}_u^-$ cover the entire item space. %
As a consequence of uniform sampling, %
the item ranking that follows is repeatedly confined to %
negative instances coming from dense data regions. %
Proposition~\ref{prop:f} formalises this observation. %
\begin{proposition}\label{prop:f}
    Two-stage -- i.e., sampling followed by ranking -- NS methods that employ uniform sampling as their foundation cannot guarantee maximum item diversity measured in the latent space. %
\end{proposition}

When the candidate set $\accentset{\star}{I}_u^-$ exhibits low diversity, the recommender repeatedly compares positive items to highly similar negatives. %
This results in redundant gradient updates and can cause the model to overfit to small regions of the item space. %
Consequently, user embeddings become less generalisable and the model may fail to sufficiently penalise items from under-sampled or sparse data regions~\cite{hong2024diversified}. Ultimately, this undermines the model's ability to capture the global structure of the item space and reduces the robustness of the ranking that it outputs. It is thus essential that negative samples used during training are sufficiently diverse to stabilise learning and promote generalisation. %

\subsection{Empirical Analysis}%

To demonstrate the limitations of uniform negative sampling and its adverse impact on instance diversity, we construct a toy example by projecting item embeddings into a two-dimensional space. %
We then use the resulting representation -- shown in Figure~\ref{fig:two_sampling} -- to compare uniform and $k$-DPP sampling. %

Specifically, Figure~\ref{fig:random_sampling} demonstrates that uniform sampling tends to select negative items that are densely concentrated within a few clusters. %
This is expected as such an approach ignores the underlying structure of the embedding space. 
In contrast, sampling based on $k$-DPP, shown in Figure~\ref{fig:dpp_sampling}, explicitly encourages diversity by penalising selection of similar items. Consequently, the sampled negatives are distributed more evenly across the latent space. %

These observations further support our theoretical analysis showing that uniform sampling fails to optimise for diversity, %
outputting instances %
drawn from selected few (dense) regions of the item space. %
In contrast, $k$-DPP sampling captures the broader structure of the embedding space and delivers more representative and diverse negatives. %
Such diversity-aware selection can improve model training stability by avoiding repeated pairing of similar negative items with different positives  
as well as enhance model expressiveness.

However, applying $k$-DPP sampling directly over a user's entire negative item set $I_u^-$ is computationally prohibitive; while it can achieve maximal diversity, its complexity is $\mathcal{O}(|I_u^{-}|^3)$. Repeating this process for every positive item of each user during model training is thus infeasible at scale. 
This limitation motivates the need for a new NS strategy that retains the diversity offered by $k$-DPP while remaining computationally efficient for large-scale datasets. %

\section{Diverse Negative Sampling}%

Our proposed method -- DivNS -- is a %
computationally efficient negative sampling strategy for recommender systems that targets item diversity. %
Existing samplers often focus on informativeness of individual negatives but overlook their diversity, thus restricting the recommender's exposure to the broader item embedding space and curtailing its generalisation ability. %
DivNS addresses these limitations by introducing diversity criteria directly into the sampling pipeline without requiring an additional model for scoring candidate negatives. An overview of our framework is shown in Figure~\ref{fig:flow-chart}. %

DivNS comprises three key modules: (\S\ref{step:1})~\emph{cache construction}, (\S\ref{sec:dpp})~\emph{diversity-augmented sampling} and (\S\ref{step:3})~\emph{synthetic negatives generation}. 
The first module maintains a cache for each user containing previously scored negative items that are highly informative but not yet ranked as the top one (i.e., the \emph{hard} negative); these instances are often ignored by NS methods despite being highly valuable. %
The second module applies a tailored $k$-DPP sampling to select a diverse subset of negatives from each user's cache while also ensuring that these instances differ sufficiently from the user's latest hard negatives; %
by limiting the sampling scope to the cached items, DivNS remains computationally efficient. %
Finally, the third module synthesises the diverse negatives and the hard negatives into a synthetic training dataset. %

\subsection{Cache Construction}\label{step:1}

Cache construction begins with two-stage sampling -- captured by \step{1}, which is indicated by a dashed box, in Figure~\ref{fig:flow-chart} -- at the start of each training epoch $t \in [0,\ldots,e]$. %
For each positive item $i^+$ of user $u$, we first uniformly sample a candidate set of negatives $\accentset{\star}{I}_u^-$ from the user's unobserved items $I_u^{-}$. We then predict the preference score $\hat{y}(u,i^-)$ for all $i^- \in \accentset{\star}{I}_u^-$ and select the item with the highest score as the \emph{hard negative}, i.e., $\accentset{\star}{i}_u^- = \arg\max_{i^- \in \accentset{\star}{I}_u^-} \hat{y}(u,i^-)$. %
Unlike standard samplers, which immediately include $\accentset{\star}{i}_u^-$ in negative training data, we defer its use. %
Instead, we collect all such hard negatives across a user's positive interactions into a dedicated hard negatives set $H_u$, where $|H_u| = |I_u^{+}|$; %
$H_u$ constructed at epoch $t$ is denoted by $H_u^{(t)}$.

In the second stage -- \step{2} in Figure~\ref{fig:flow-chart} -- %
we extract the top-$m$ ranked items from $\accentset{\star}{I}_u^- \setminus \{\accentset{\star}{i}_u^-\}$ and store them in a user-specific cache $C_u$. %
Here, $m$, called \emph{cache ratio}, controls the number of items cached for each positive interaction, with $m\ll n=|\accentset{\star}{I}_u^-|$; %
the cache size is thus $|C_u|=m\times|I_u^{+}|$. %
As the recommender parameters get updated throughout training, $C_u$ needs to be refreshed at every epoch to retain up-to-date informative negatives; %
we denote the user-specific cache at epoch $t$ as $C_u^{(t)}$. 

The items in $C_u$ %
are informative given their high predicted scores. %
Nonetheless, they are typically discarded by NS methods because they are not the highest-ranked items in $\accentset{\star}{I}_u^-$. %
Since multiple informative negatives often reside in each $\accentset{\star}{I}_u^-$~\cite{ding2020simplify}, we retain these valuable instances in $C_u$ for model training. %
We deliberately select only the top-$m$ items from $\accentset{\star}{I}_u^-$, instead of using all $n$ candidates, for two reasons. First, only a few of the top-ranked negatives in $\accentset{\star}{I}_u^-$ are informative, hence using the full set may introduce excessive noise~\cite{wu2021rethinking}. Second, the computational cost of $k$-DPP sampling in the next stage scales with the size of $C_u$; %
as this value depends on $m$, keeping the cache ratio small improves efficiency. %

\subsection{Diversity-Augmented Sampling}\label{sec:dpp}

After constructing the cache $C_u^{(t)}$, we begin the next epoch $t+1$ -- \step{3} in Figure~\ref{fig:flow-chart} -- by selecting new hard negatives for each positive item $i^+$ of user $u$, thus forming the hard negatives set $H_u^{(t+1)}$. %
Then, we sample the set of diverse negatives $D_u^{(t+1)}$ from $C_u^{(t)}$ using $k$-DPP to this end -- %
a process that inherently favours subsets of items with diverse latent space embeddings. %
Since both $D_u^{(t+1)}$ and $H_u^{(t+1)}$ are used later to enrich the training data, we further increase data diversity by encouraging $D_u^{(t+1)}$ to be dissimilar from items in $H_u^{(t+1)}$. We achieve this by incorporating a penalty into the kernel used by $k$-DPP, %
thus suppressing %
selection of items that are similar to those in $H_u^{(t+1)}$.

As any two items $i, j \in \mathcal{I}$ can be represented by embedding vectors $\mathbf{v}_i, \mathbf{v}_j \in \mathbb{R}^d$, we measure their similarity with a kernel $\kappa : \mathbb{R}^d \times \mathbb{R}^d \mapsto \mathbb{R}$ %
defined as $\kappa(\mathbf{v}_i, \mathbf{v}_j) = \mathbf{v}_i^\top \mathbf{v}_j$. %
$k$-DPP then specifies a probability distribution over all the possible subsets $D \subseteq C_u^{(t)}$ as: %
\begin{equation}\label{eq:dpp-prob}
\mathcal{P}(D) \propto \det(\mathbf{K}_D)~\text{,} %
\end{equation}
where $\mathbf{K}_D = (k_{ij})$ is the kernel matrix composed of $k_{ij} = \kappa(\mathbf{v}_i, \mathbf{v}_j)$ for $i, j \in D$. %
For efficiency, we compute $\mathbf{K}_{C_u^{(t)}}$ once and get any $\mathbf{K}_D$ for free by extracting its sub-matrix indexed by the items in $D$. %

$k$-DPP naturally favours sets of items that are dissimilar, i.e., have low pairwise similarity, given the determinant structure. %
To further ensure that the item set sampled by $k$-DPP from $C_u^{(t)}$ is also dissimilar from the hard negatives $H_u^{(t+1)}$, we augment the kernel function $\kappa$ %
with a \emph{diversity} criterion as so: %
$$
   \widetilde{\kappa}(\mathbf{v}_i, \mathbf{v}_j)
    = q_i q_j \kappa(\mathbf{v}_i, \mathbf{v}_j) ~\text{,} %
$$
where $q_i, q_j \in [0,1]$ are 
penalty factors for items $i$ and $j$ respectively; %
they are defined as: %
\begin{equation}\label{eq:penalty_factor}
      \begin{aligned}
    q_a
      & = \frac{1}{|H_u^{(t+1)}|}\sum_{b \in H_u^{(t+1)}} \left( 1 - \cos(\theta_{ab}) \right) \\
      & = 1 - \frac{1}{|H_u^{(t+1)}|}\sum_{b \in H_u^{(t+1)}} %
      \mathbf{v}_a^\top \mathbf{v}_b ~\text{,}
      \end{aligned}
\end{equation}
thus %
down-weighing the contribution (hence marginal probability) of items similar to those in $H_u^{(t+1)}$ %

By using $k$-DPP with $k = |H_u^{(t+1)}|$, rather than vanilla DPP, we %
ensure that our set $D_u^{(t+1)}$ sampled from $C_u^{(t)}$ has the same cardinality as the hard negatives set $H_u^{(t+1)}$. %
To avoid any overlap between $D_u^{(t+1)}$ and $H_u^{(t+1)}$, we explicitly remove all the items from the cache $C_u^{(t)}$ that are already present in $H_u^{(t+1)}$. %
For computational efficiency, we adopt a greedy approximation of exact $k$-DPP sampling based on Cholesky updates~\cite{gillenwater2012near}. 
Consequently, our bespoke sampling approach ensures that the resulting set $D_u^{(t+1)}$ is \emph{diverse within itself} because of how $k$-DPP works as well as \emph{distinct from the hard negatives set $H_u^{(t+1)}$} thanks to our custom kernel function $\widetilde{\kappa}$. %
$D_u^{(t+1)}$ thus provides high-quality diverse items that complement the hard negatives, %
allowing them to enrich the training data of each user as we explain in the \emph{synthetic negatives generation} step. %

\paragraph{Theoretical Complexity Analysis}
The computational complexity of our bespoke $k$-DPP sampler is dominated by %
construction of the kernel matrix $\mathbf{K_{C_u}}$, %
which %
requires $\mathcal{O}({|C_u|}^2\times d)$ calculations, where $d$ is the embedding dimensionality. %
Additionally, sampling the subset $D_u$ of size $|D_u|=|I_u^{+}|$ from $C_u$ costs $\mathcal{O}({|C_u|}^2 \times |I_u^{+}|)$ when using a greedy MAP-based approximation of $k$-DPP. %
Overall, our sampler runs in %
$\mathcal{O}({|C_u|}^2 \times (|I_u^{+}| + d) )$. 
Since $|I_u^{+}|$ is small because of data sparsity and $d$ is pre-defined, the efficiency mainly depends on $|C_u|=m \times |I_u^{+}|$. %
We thus %
choose to %
keep the cache ratio $m$ small.
Further details of this complexity analysis are given in Appendix~\ref{apx:tc}. %

\subsection{Synthetic Negatives Generation}\label{step:3}

We update the hard negative $i^H \in H_u$ paired with each positive item $i^+ \in I_u^{+}$ by mixing it with a random diverse negative $i^D \in D_u$ -- \step{5} in Figure~\ref{fig:flow-chart}. %
This synthetic instance is defined as: %
\begin{equation}\label{eq:mixup}
    \mathbf{v}_{i^\ominus} = \lambda\,\mathbf{v}_{i^H} + (1-\lambda)\,\mathbf{v}_{i^D}~\text{,}
\end{equation}
where $\lambda\in[0,1]$ is the mixing parameter. %
This procedure is inspired by mixup~\cite{zhang2017mixup}, which creates new instances via linear interpolation. %

New training data composed of $(u, i^+, i^\ominus)$ triplets are then used with the BPR loss %
to update the model parameters via gradient descent -- \step{6} -- and refresh the caches -- \step{7}. %
Since every $i^\ominus$ lies between hard and diverse negatives in the latent space, the model is able to better generalise and avoid overfitting to a narrow selection of similar items. %
We then proceed to the next epoch and restart the process from \step{3}. %
Taken together, our framework combines dynamic mining of hard negatives with sampling based on $k$-DPP and mixup synthesis to produce a rich and diverse set of negative instances for model training. %

\begin{algorithm}[b]
\caption{Two-step optimisation implementation of DivNS.}
\label{alg:diverse-neg-sampling}
\small
\begin{algorithmic}[1]
\REQUIRE Training set $\mathcal{Y}=\{(u,i) \; | \; u\in \mathcal{U}, i\in \mathcal{I}\}$; %
model scoring function $\hat{y}(\cdot, \cdot)$; %
(initial) model parameters $\Theta$; number of epochs $t_{\max}$; candidate set size $n$; cache ratio $m$.
\ENSURE Model parameters $\Theta$ are updated.
\FOR{$t = 0$; $t = t + 1$; \textbf{until} $t < t_{\max}$}
  \STATE \textit{\color{gray!95}// Inner diverse sampling step}
  \FOR{$u\in \mathcal{U}$}
    \STATE $C_u^{(t)} \gets \emptyset$, \, $H_u^{(t)} \gets \emptyset, \, I_u^+ = \{i \; | \; (u,i) \in \mathcal{Y}, u=u\}, \, I_u^- = \mathcal{I} \setminus I_u^+$\label{alg:ref1}%
    \FOR{$i^+ \in I_u^{+}$}
      \STATE $\accentset{\star}{I}_u^- \gets \texttt{RandomChoice}(I_u^-, \, n)$ %
\COMMENT{\textit{\color{gray!95}{Sample without replacement}}}
      \STATE $ \accentset{\star}{i}_u^- \gets \arg\max_{i^- \in \accentset{\star}{I}_u^-}{\hat{y}(u,i^-)}$
      \STATE $H_u^{(t)} \gets H_u^{(t)} \cup \{ \accentset{\star}{i}_u^- \}$
      \STATE $C_u^{(t)} \gets C_u^{(t)} \cup \left\{\arg\max_{i^- \in \accentset{\star}{I}_u^- \setminus \{\accentset{\star}{i}_u^-\}}{\hat{y}(u,i^-)}\right\}_1^m$ %
    \ENDFOR \label{alg:ref2}
    \IF{$t > 0$}
      \STATE $D_u^{(t)} \gets k\texttt{-DPP}\bigl(C_u^{(t-1)} \setminus H_u^{(t)}, \, |H_u^{(t)}|\bigr)$ \label{alg:ref3}
      \STATE $I_u^\ominus \gets \emptyset$
      \FOR{$i^H \in H_u^{(t)}$}
        \STATE $i^D \gets \texttt{RandomChoice}(D_u^{(t)}, \, 1)$
        \STATE $i^\ominus \gets \texttt{mixup}(i^H, \, i^D)$ \COMMENT{\textit{\color{gray!95}{Equation~\ref{eq:mixup}}}} %
        \STATE $I_u^\ominus \gets I_u^\ominus \cup \{i^\ominus\}$, \,
        $D_u^{(t)} \gets D_u^{(t)} \setminus \{i^D\}$
      \ENDFOR \label{alg:ref4}
    \ELSE
      \STATE $I_u^\ominus \gets H_u^{(t)}$
    \ENDIF
  \ENDFOR
  \STATE \textit{\color{gray!95}// Outer optimisation step}%
  \STATE Backpropagate $\ell_{\mathit{BPR}}$ computed over $(u, i^+, i^\ominus)$ to update $\Theta$ \label{alg:ref5} %
\ENDFOR
\end{algorithmic}
\end{algorithm}

\subsection{Two-Step Optimisation Implementation}

The implementation of DivNS has two main components -- inner diverse sampling and outer optimisation -- that are run in each epoch $t$ as shown in Algorithm~\ref{alg:diverse-neg-sampling}.  %
The inner sampling step begins by collecting hard negatives and caching informative candidates (Lines~\ref{alg:ref1}--\ref{alg:ref2}). %
Specifically, for every positive item $i^+$ of each user $u$, we first uniformly sample the set of unobserved items $\accentset{\star}{I}_u^-$; then, we pick the top-scoring instance as the hard negative, storing it in $H_u^{(t)}$; lastly, we add the remaining top-$m$ samples to the user's cache $C_u^{(t)}$.  %
Throughout the epoch, $H_u^{(t)}$ accumulates hard negatives and $C_u^{(t)}$ is gradually expanded for each user $u$. %

The inner sampling step continues (Lines~\ref{alg:ref3}--\ref{alg:ref4}) %
by excluding the current hard negatives $H_u^{(t)}$ from %
the cache $C_u^{(t-1)}$ built in the previous epoch; %
we then sample $k = |H_u^{(t)}|$ items from this set with $k$-DPP %
to form the diverse set $D_u^{(t)}$. %
Each hard negative in $H_u^{(t)}$ is next paired with a random item from $D_u^{(t)}$; their mixing gives the synthetic negative $i^\ominus$, which is stored in $I_u^\ominus$, thus concluding the inner sampling step. %
The outer optimisation step (Line~\ref{alg:ref5}) computes %
the BPR loss over the $(u, i^+, i^\ominus)$ triplets to update the model. %

\section{Experiments}\label{sec:exp}
We conduct extensive experiments on four popular datasets with nine representative negative samplers as the baselines to demonstrate the effectiveness of DivNS. %

\subsection{Setup}\label{sec:exp-setup}

\paragraph{Negative Samplers}
We compare DivNS to nine NS approaches; the first two are \emph{static} and the remaining seven are \emph{hard} samplers: %
\begin{description}%
    \item [Random Negative Sampling (RNS)] applies the uniform distribution to randomly sample negative items~\cite{rendle2012bpr}. 
    \item [Popularity-based NS (PNS)] assigns higher probability to popular items that the user has not yet interacted with~\cite{chen2017sampling}.
    \item [Dynamic NS (DNS)] uses two-stage sampling that first uniformly samples a candidate set of negative items and then selects the one with the highest predicted score~\cite{zhang2013optimizing}.
    \item [Importance Resampling (AdaSIR)] uses historical hard samples with importance resampling to maintain candidate sets~\cite{chen2022learning}.
    \item [DNS(M, N)] extends DNS by using M to control sampling hardness and N to determine the size of the
    candidate set~\cite{shi2023theories}.
    \item [MixGCF] interpolates negatives by mixing them with neighbouring instances %
    as well as positive items~\cite{huang2021mixgcf}.
    \item [Disentangled NS (DENS)] disentangles irrelevant and relevant\break factors of items, contrasting them to select the best negatives~\cite{lai2023disentangled}. %
    \item [Adaptive Hardness NS (AHNS)] adaptively picks negative items with varying hardness levels for each positive item~\cite{lai2024adaptive}.  %
    \item [Stochastic Contrastive NS (SCONE)] leverages stochastic positive injection to generate hard negatives for graph-based recommenders~\cite{lee2025scone}. %
\end{description}

We do not experiment with samplers like SRNS~\cite{ding2020simplify} or MCNS~\cite{yang2020understanding} as research has shown their subpar performance~\cite{huang2021mixgcf,chen2023revisiting,lai2023disentangled,lai2024adaptive}.  %
We integrate samplers with two representative recommenders: Matrix Factorisation (MF) and graph-based Light Graph Convolution Network (LightGCN)~\cite{he2020lightgcn}. %
Among our nine baselines, MixGCF and SCONE are by design only compatible with graph-based recommenders; the other methods work with both recommenders. %

\begin{table*}[t]
    \footnotesize
    \setlength{\tabcolsep}{2.55pt}%
    \caption{Recommendation utility -- measured using NDCG and Recall based on top-10 and top-20 ranking -- of negative samplers paired with two recommenders: MF and LightGCN. The best results are in bold and the second best results are underlined. Relative improvement -- RI -- of DivNS over the best baseline is %
    marked with $\ast$ whenever it is statistically significant ($p<0.05$). %
    }
    \begin{tabular}{@{}p{0.20cm}lrrrrrrrrrrrrrrrr@{}}
    \toprule
     &  & \multicolumn{8}{c}{Top-10} & \multicolumn{8}{c}{Top-20}  \\
     \cmidrule(lr){3-10}\cmidrule(lr){11-18} 
     &  & \multicolumn{2}{c}{Amazon Beauty} & \multicolumn{2}{c}{ML 1M} & \multicolumn{2}{c}{Pinterest} & \multicolumn{2}{c}{Yelp 2022} & \multicolumn{2}{c}{Amazon Beauty} & \multicolumn{2}{c}{ML 1M} & \multicolumn{2}{c}{Pinterest} & \multicolumn{2}{c}{Yelp 2022} \\
     \cmidrule(lr){3-4} \cmidrule(lr){5-6} \cmidrule(lr){7-8}
     \cmidrule(lr){9-10} \cmidrule(lr){11-12} \cmidrule(lr){13-14}
     \cmidrule(lr){15-16} \cmidrule(lr){17-18} 
     & Sampler & NDCG & Recall & NDCG & Recall & NDCG & Recall & NDCG & Recall & NDCG & Recall & NDCG & Recall & NDCG & Recall & NDCG & Recall 
     \\ \midrule
     \multirow{9}{*}{\rotatebox[origin=c]{90}{\parbox[c]{2cm}{\centering MF}}} 
     & RNS & 0.0275 & 0.0471 & 0.2624 & 0.1948 & 0.0514 & 0.0812 & 0.0408 & 0.0476 & 0.0304 & 0.0577 & 0.3009 & 0.2236 & 0.0766 & 0.1160 & 0.0527 & 0.0613   \\
     & PNS & 0.0266 & 0.0466 & 0.2550 & 0.1837 & 0.0465 & 0.0783 & 0.0393 & 0.0477 & 0.0303 & 0.0572 & 0.3010 & 0.2235 & 0.0640 & 0.0983 & 0.0517 & 0.0598 \\
     & DNS & 0.0310 & \underline{0.0495} & 0.2712 & 0.2018 & \underline{0.0688} & \underline{0.0840} & 0.0459 & 0.0556 & \underline{0.0354} & \underline{0.0703} & \underline{0.3328} & \underline{0.2289} & \underline{0.0814} & \underline{0.1263} & 0.0582 & 0.0697 \\
     & AdaSIR & 0.0296 & 0.0487 & 0.2674 & 0.1978 & 0.0654 & 0.0795 & 0.0448 & 0.0527 & 0.0322 & 0.0627 & 0.3114 & 0.2257 & 0.7061 & 0.1135 & 0.0563 & 0.0663   \\
     & DNS(M,~N) & \underline{0.0314} & 0.0493 & 0.2705 & 0.2016 & 0.0674 & 0.0826 & 0.0457 & 0.0535 & 0.0339 & 0.0645 & 0.3188 & 0.2269 & 0.0782 & 0.1178 & 0.0573 & 0.0686  \\ 
     & DENS & 0.0303 & 0.0485 & \underline{0.2713} & \underline{0.2024} & 0.0656 & 0.0813 & \underline{0.0466} & \underline{0.0534} & 0.0352 & 0.0699 & 0.3325 & 0.2286 & 0.0736 & 0.1120 & \underline{0.0592} & \underline{0.0703}  \\
     & AHNS & 0.0297 & 0.0483 & 0.2547 & 0.1949 & 0.0653 & 0.0811 & 0.0452 & 0.0524 & 0.0332 & 0.0631 & 0.2959 & 0.2154 & 0.0706 & 0.1052 & 0.0568 & 0.0675  \\
     \cmidrule(lr){2-18}
     & DivNS & \textbf{0.0327} & \textbf{0.0518} & \textbf{0.2812} & \textbf{0.2125} & \textbf{0.0733} & \textbf{0.0882} & \textbf{0.0486} & \textbf{0.0551} & \textbf{0.0385} & \textbf{0.0758} & \textbf{0.3397} & \textbf{0.2393} & \textbf{0.0883} & \textbf{0.1356} & \textbf{0.0613} & \textbf{0.0727} \\ \cmidrule(lr){2-18}
     & RI & 4.14\%$^\ast$ & 4.65\%$^\ast$ & 3.65\%$^\ast$ & 5.00\%$^\ast$ & 6.54\%$^\ast$ & 5.00\%$^\ast$ & 4.29\%$^\ast$ & 3.18\%$^\ast$ & 8.76\%$^\ast$ & 7.82\%$^\ast$ & 2.07\% & 4.54\%$^\ast$ & 8.48\%$^\ast$ & 7.36\%$^\ast$ & 3.38\%$^\ast$ & 3.41\%$^\ast$ \\
     \midrule
    \multirow{10}{*}{\rotatebox[origin=c]{90}{\parbox[c]{2.0cm}{\centering LightGCN}}} 
     & RNS & 0.0303 & 0.0624 & 0.2501 & 0.1823 & 0.0745 & 0.0994 & 0.0516 & 0.0598 & 0.0398 & 0.0802 & 0.3225 & 0.2304 & 0.0951 & 0.1405 & 0.0666 & 0.0753 \\
     & PNS & 0.0301 & 0.0586 & 0.2375 & 0.1792 & 0.0646 & 0.0948 & 0.0473 & 0.0575 & 0.0393 & 0.0797 & 0.3152 & 0.2288 & 0.0853 & 0.1273 & 0.0637 & 0.0728 \\
     & DNS & 0.0322 & \underline{0.0674} & \underline{0.2762} & 0.1895 & 0.0728 & 0.1077 & \underline{0.0587} & \underline{0.0677} & 0.0427 & 0.0838 & 0.3437 & 0.2412 & 0.0986 & 0.1449 & \underline{0.0755} & \underline{0.0862} \\
     & AdaSIR & 0.0310 & 0.0642 & 0.2540 & 0.1804 & 0.0692 & 0.0914 & 0.0540 & 0.0656 & 0.0404 & 0.0823 & 0.3220 & 0.2306 & 0.0931 & 0.1237 & 0.0706 & 0.0832 \\
     & DNS(M,~N) & 0.0315 & 0.0660 & 0.2560 & 0.1935 & 0.0727 & 0.1050 & 0.0574 & 0.0663 & 0.0408 & 0.0825 & 0.3313 & 0.2396 & 0.0944 & 0.1389 & 0.0736 & 0.0849 \\ 
     & MixGCF & \underline{0.0323} & 0.0672 & \underline{0.2762} & \underline{0.1976} & 0.0801 & 0.1122 & 0.0585 & 0.0656 & 0.0426 & \underline{0.0841} & \underline{0.3442} & \underline{0.2414} & \underline{0.1010} & 0.1482 & 0.0755 & 0.0854 \\ 
     & DENS & 0.0321 & 0.0626 & 0.2478 & 0.1892 & 0.0706 & 0.0963 & 0.0567 & 0.0663 & 0.0397 & 0.0800 & 0.3223 & 0.2319 & 0.0800 & 0.1183 & 0.0740 & 0.0850 \\
     & AHNS & 0.0310 & 0.0597 & 0.2418 & 0.1790 & 0.0703 & 0.0855 & 0.0555 & 0.0649 & 0.0392 & 0.0791 & 0.3045 & 0.2287 & 0.0912 & 0.1185 & 0.0734 & 0.0812 \\ 
     & SCONE & 0.0321 & 0.0670 & 0.2758 & 0.1919 & \underline{0.0814} & \underline{0.1131} & 0.0584 & 0.0668 & \underline{0.0428} & 0.0838 & 0.3436 & 0.2411 & 0.1003 & \underline{0.1485} & 0.0752 & 0.0861 \\
     \cmidrule(lr){2-18}
     & DivNS & \textbf{0.0340} & \textbf{0.0693} & \textbf{0.2827} & \textbf{0.2053} & \textbf{0.0827} & \textbf{0.1170} & \textbf{0.0603} & \textbf{0.0690} & \textbf{0.0440} & \textbf{0.0859} & \textbf{0.3485} & \textbf{0.2486} & \textbf{0.1054} & \textbf{0.1536} & \textbf{0.0788} & \textbf{0.0886} \\ \cmidrule(lr){2-18}
     & RI & 5.26\%$^\ast$ & 2.82\%$^\ast$ & 2.35\%$^\ast$ & 3.90\%$^\ast$ & 1.60\% & 3.45\%$^\ast$ & 2.73\%$^\ast$ & 1.92\% & 2.80\%$^\ast$ & 2.14\% & 1.25\% & 2.98\%$^\ast$ & 4.36\%$^\ast$ & 3.43\%$^\ast$ & 4.37\%$^\ast$ & 2.78\%$^\ast$ \\
    \bottomrule
    \end{tabular}
    \label{tab:results}
    \end{table*}

\paragraph{Datasets}
We use four benchmark datasets: Amazon Beauty, MovieLens (ML) 1M, Pinterest and Yelp 2022; their statistics are summarised in Appendix~\ref{app:setup}. %
Our selection covers sparse and dense datasets of varying size~\cite{chin2022datasets}. %
For Yelp, we use its latest release -- Yelp 2022 -- which is also the largest. %
Item ratings are binarised to form implicit feedback data that indicate whether a user has interacted with an item. 
Since all the baselines and recommenders are non-sequential, we follow the standard fixed-ratio dataset split widely used in negative sampling studies and partition each dataset into training, validation and test sets using a 70--10--20\% split. 
Most sampling methods are designed and evaluated under this setup~\cite{wang2019neural,yang2020understanding,huang2021mixgcf,lee2025scone}. %

\paragraph{Evaluation}%
We use two metrics: NDCG@$p$ and Recall@$p$. %
Following common practice, we choose $p\in \{10, 20\}$~\cite{huang2021mixgcf,chen2023revisiting,chen2023fairly,lai2023disentangled,lai2024adaptive}. During evaluation, a recommender outputs a ranking of all items except those already in a user's training set; we therefore do not use sampling during evaluation in line with best practice~\cite{krichene2020sampled}. %

\paragraph{Hyperparameters}%
We implement all recommenders and samplers, including DivNS, in PyTorch with the Adam optimiser. For each method, we conduct grid search to find the optimal hyperparameters, including the learning rate $lr$, regularisation weight $l_2$, candidate size $n$, cache ratio $m$ and mixing coefficient $\lambda$. %
We fix the embedding size to $d = 64$ following common practice~\cite{huang2021mixgcf, lai2023disentangled, lai2024adaptive, lee2025scone}. %
Additional details about our setup are provided in Appendix~\ref{app:setup}. %
Our code is available at: %
\href{https://github.com/xuanxuanxuan-git/DivNS}{github.com/xuanxuanxuan-git/DivNS}.

\subsection{Results}\label{sec:result}

Each experiment is repeated five times with a different random seed; Table~\ref{tab:results} reports their average result.
We find that, across all four datasets, DivNS consistently outperforms all the baselines by a substantial margin.
This improvement stems from two key design choices. 
First, by constructing user-specific caches that contain top-ranked negatives, more informative instances are used, thus reducing the amount of noise. %
Second, by applying diversity-augmented $k$-DPP sampling, we further allow the recommender to explore a broader item space and learn a more precise decision boundary for each user, thereby improving model generalisation. %

The performance gain from using DivNS is more pronounced for MF than LightGCN when compared to all the baselines. %
We speculate that this is because MixGCF is an effective negative sampler (recall that it is incompatible with MF), which also has a component that mixes negatives. %
Distinct from our method, MixGCF synthesises a negative instance with its neighbouring negatives in a graph;
it therefore only implicitly introduces diverse item sets to the model. %
In contrast, our method directly optimises for diversity, helping it to outperform MixGCF. %
Moreover, none of the existing MF-compatible samplers account for the diversity of sampled negative data, giving DivNS a significant advantage when it is paired with MF. %

DivNS extends the standard DNS approach by not only identifying hard negatives but also synthesising them with other diverse and informative negative items. The consistent improvement achieved by DivNS over DNS highlights the critical role of diversity in negative sampling as we elaborate in the following section. %

\paragraph{Runtime}%
Taking LightGCN as an example, the average training times per epoch for RNS, DNS, MixGCF, DENS and DivNS on Amazon Beauty are approximately 5, 20, 26, 29 and 31 seconds respectively; 
runtime results for the other datasets are listed in Appendix~\ref{app:add-results}. %
We find that, although DivNS requires slightly more compute due to $k$-DPP sampling, setting the cache ratio $m$ to a small value (as discussed in Section~\ref{sec:dpp}) allows to keep its training time on a par with other hard negative samplers, making it competitive. %

\begin{figure}[b]
    \centering
    \includegraphics[width=0.8\linewidth]{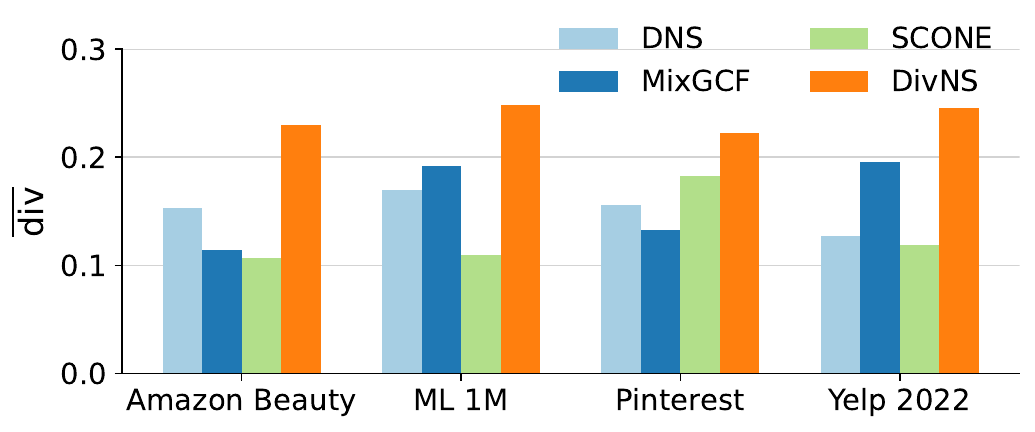}
    \caption{Diversity of sampled negatives %
    for the top four performing samplers deployed with LightGCN.}%
    \label{fig:different-div}
\end{figure}

\paragraph{Sampled Negatives Diversity}%
To verify that DivNS enhances sampling diversity, we measure this property for the negative items used during model training. Let ${I_u^\ominus}^{(t)}$ denote the set of negative training data sampled by DivNS for user $u$ in epoch $t$. The average diversity across all users and epochs is %
$    
    \overline{\mathit{div}} = \frac{1}{t_{\max}}\sum_{u\in \mathcal{U}} \sum_{t=0}^{t_{\max}-1}{\mathit{div}({I_u^\ominus}^{(t)})}%
$,
where $\mathit{div}(\cdot)$ is the diversity metric defined in Equation~\ref{eq:sim-metric}. Figure~\ref{fig:different-div} compares $\overline{\mathit{div}}$ of DivNS to top baselines for LightGCN; the remaining results are listed in Appendix~\ref{app:add-results}. %
We find that DivNS consistently produces more diverse negative samples than all the baselines, confirming the effectiveness of our diversity-aware approach. %

\subsection{DivNS Analysis and Ablation Study}\label{sec:ablation}

Next, we conduct an ablation study to gauge the impact of each DivNS module. %
Given the sequential structure of our approach, %
removing the first module (cache construction) disables the entire pipeline. %
Our evaluation therefore relies on: %
(1)~removing the full pipeline; %
(2)~replacing our diversity-augmented $k$-DPP sampling (module two) with random and vanilla $k$-DPP sampling; and %
(3)~tweaking our synthetic negatives generation (module three) by varying the mixing coefficient $\lambda$ (Equation~\ref{eq:mixup}).  %
We also examine the influence of a key DivNS hyperparameter: the cache ratio $m$. %
To ensure fair comparison, all the experiments follow a setup that is identical to the one described in Section~\ref{sec:exp-setup}. %

\begin{table}[b!]
    \centering
    \footnotesize
    \setlength{\tabcolsep}{2.55pt}%
    \caption{Performance -- NDCG@20 and Recall@20 -- of DivNS with (\cmark) and without (\xmark) its all three modules. %
    }
    \begin{tabular}{@{}lrrrrrrrrr@{}}%
    \toprule
     & & \multicolumn{2}{c}{Amazon Beauty} & \multicolumn{2}{c}{ML 1M} & \multicolumn{2}{c}{Pinterest} & \multicolumn{2}{c}{Yelp 2022}  \\
     \cmidrule(lr){3-4} \cmidrule(lr){5-6} \cmidrule(lr){7-8}
     \cmidrule(lr){9-10}  
     & & NDCG & Recall & NDCG & Recall & NDCG & Recall & NDCG & Recall \\ \midrule
     \multirow{2}{*}{MF} & \cmark & \textbf{0.0385} & \textbf{0.0758} & \textbf{0.3397} & \textbf{0.2393} & \textbf{0.0883} & \textbf{0.1356} & \textbf{0.0613} & \textbf{0.0727} \\
     & \xmark  & 0.0354 & 0.0703 & 0.3328 & 0.2289 & 0.0814 & 0.1263 & 0.0582 & 0.0697 \\
     \midrule
     \multirow{2}{*}{LightGCN} & \cmark & \textbf{0.0440} & \textbf{0.0859} & \textbf{0.3485} & \textbf{0.2486} & \textbf{0.1054} & \textbf{0.1536} & \textbf{0.0788} & \textbf{0.0886} \\
     & \xmark  & 0.0427 & 0.0838 & 0.3437 & 0.2412 & 0.0986 & 0.1449 & 0.0755 & 0.0862 \\
    \bottomrule
    \end{tabular}
    \label{tab:ablation-no-syn}
\end{table}

\paragraph{Disable All Modules: Using Diverse Negatives}%
To verify the impact of using diverse negatives in recommender training, we disable DivNS entirely, %
effectively reverting the sampling step to standard DNS (i.e., using only the top-ranked, hard, negative for each positive). %
Results in Table~\ref{tab:ablation-no-syn} show performance decline of 5--10\% when diverse negatives are absent. %
This finding confirms the importance of diversity in negative training data and supports the core motivation behind our work (outlined earlier in Section~\ref{sec:motivation}). %

\paragraph{Swap Module Two: Diversity-Augmented Sampling}%
Since di\-ver\-si\-ty-augmented $k$-DPP is the foundation of our method, we compare its effectiveness to other sampling strategies. %
First, we experiment with vanilla $k$-DPP, in which case the negatives in $D_u$ may contain items similar to those in $H_u$, sacrificing the training data diversity. %
We also test uniform sampling, which further weakens the diversity of $D_u$. 
The exact results of these experiments are reported in Appendix~\ref{app:add-results}. %
We observe that our diversity-augmented $k$-DPP sampling consistently delivers the best performance. Vanilla $k$-DPP (which lacks our bespoke kernel) performs slightly worse, but it still outperforms uniform sampling in most cases. These findings highlight the central role of data diversity in enhancing recommender training effectiveness. %

\begin{figure}[t]
    \begin{subfigure}{0.48\linewidth}
        \includegraphics[width=\linewidth]{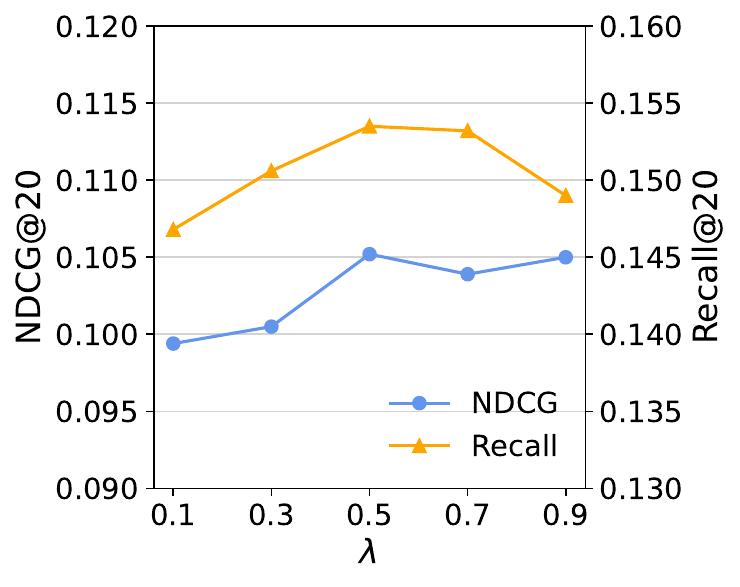}
        \caption{Pinterest.}
        \label{fig:pinterest-lambda}
    \end{subfigure}
    \hfill%
    \begin{subfigure}{0.48\linewidth}
        \includegraphics[width=\linewidth]{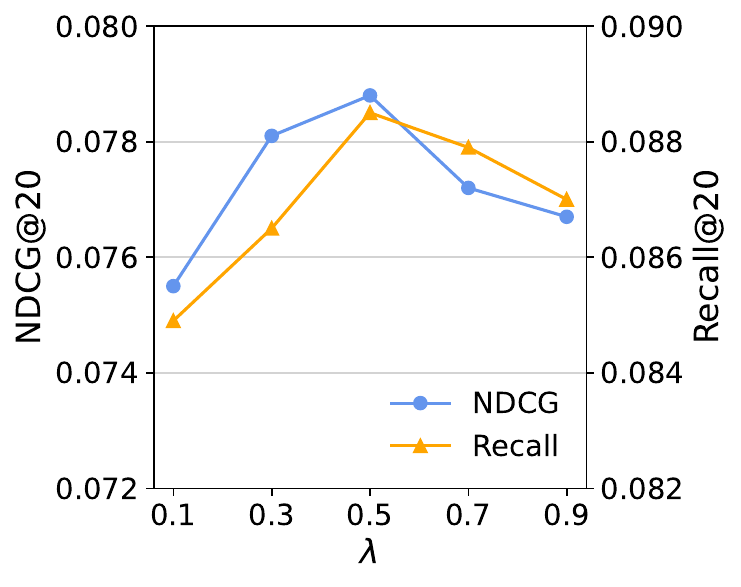}
        \caption{Yelp 2022.}
        \label{fig:yelp-lambda}
    \end{subfigure}
    \caption{Impact of synthetic negatives generation on performance -- NDCG@20 and Recall@20 -- of LightGCN
    for (\subref{fig:pinterest-lambda})~Pinterest and (\subref{fig:yelp-lambda})~Yelp 2022. Appendix~\ref{app:add-results} lists full results.} %
    \label{fig:ablation-lambda}
\end{figure}

\paragraph{Tweak Module Three: Synthetic Negatives Generation}%
We also study the impact of the mixing coefficient $\lambda$ (Equation~\ref{eq:mixup}), which controls the balance between diverse and hard negatives when synthesising negative training data. %
Smaller values of $\lambda$ increase the contribution of diverse negatives;   %
we vary $\lambda$ between $0.1$ and $0.9$ with a step size of $0.2$. %
Figure~\ref{fig:ablation-lambda} shows the results of this experiment. 
We observe that the performance improves as $\lambda$ decreases from $0.9$ to $0.7$ or $0.5$, which strengthens the contribution of diverse negatives; nonetheless, it starts to decline for $\lambda<0.5$.  %
This finding reflects the fundamental trade-off between diverse and hard %
negatives: while the former improve model generalisation, they are less informative than the latter, so over-relying on them can hinder model training (showing as lower NDCG@20 for small values of $\lambda$). %
However, confining recommender training exclusively to hard negatives also impairs its performance (evidenced by reduced NDCG@20 when $\lambda$ approaches $1$). %

\paragraph{Cache Ratio Adjustment}%
To further study the behaviour of cache construction, %
we vary the cache ratio $m \in \{1,2,4,6\}$, which affects how many top-ranked negatives from the candidate set $\accentset{\star}{I}_u^-$ are kept in the cache $C_u$. %
The exact results of these experiments are reported in Appendix~\ref{app:add-results}. %
We find that increasing $m$ up to $4$ improves the performance; %
but when $m=6$ with $n=10$, we cache more than half of the candidate negatives from $\accentset{\star}{I}_u^-$, which entails storing and using lower quality items that may be uninformative, thus degrading performance. %
Keeping $m<n/2$ appears to be the most effective. %

\section{Conclusion}
In this paper we proposed DivNS: a novel negative sampling framework for collaborative filtering that selects a diverse set of negative instances from across the latent item space. 
Our analyses revealed that existing negative samplers neglect data diversity, limiting the ability of recommenders to capture users' negative preferences. %
DivNS addressed this limitation by, first, constructing user-specific caches containing highly informative negative items that are discarded by traditional samplers. %
Then, it applied a diversity-augmented sampling strategy to produce a representative subset of negatives from the cache, ensuring that the selected items also differ from the top-ranked, hard, negatives. 
Finally, it synthesised the diverse and hard negatives to create more effective training examples. %
This process enabled the recommenders to explore a larger item space and learn more comprehensive user--item relationships. 
Experiments on four popular datasets showed that DivNS consistently outperforms state-of-the-art samplers while remaining computationally efficient.

\section*{Acknowledgements}
This research was conducted by the ARC Centre of Excellence for Automated Decision-Making and Society (CE200100005), %
funded by the Australian Government through the Australian Research Council. %
Additional support was provided by the TRUST-ME project (205121L 214991), funded by the Swiss National Science Foundation. %

\bibliographystyle{ACM-Reference-Format}
\bibliography{reference}

\appendix

\section{Determinantal Point Process}\label{app:dpp-background}%

A point process $\mathcal{P}$ on a discrete set $Z=\{1, 2, \ldots, n\}$ is a probability measure on $2^n$, i.e., the set of all the subsets of $\mathcal{Z}$. A Determinantal Point Process (DPP) is its variant that favours diverse subsets. %
Formally, $\mathcal{P}$ is a DPP when there exists a real, positive semidefinite kernel matrix $\mathbf{K}\in\mathbb{R}^n \times \mathbb{R}^n$ such that for any subset $Y \subseteq Z$ the probability of selecting $Y$ is: 
\begin{equation*}
    \mathcal{P}(Y)\propto \det (\mathbf{K}_Y)~\text{,}
\end{equation*}
where $\mathbf{K}_Y$ denotes the sub-matrix of $\mathbf{K}$ indexed by the elements in $Y$~\cite{kulesza2012determinantal}. 

Determinants have an intuitive geometric interpretation. The determinant $\det(\mathbf{K}_Y)$ measures the volume spanned by the feature vectors of items in $Y$. A larger volume indicates that these feature vectors are more orthogonal, hence items in $Y$ are more diverse. This explains why DPPs naturally favour diverse subsets~\cite{kulesza2012determinantal}.

One key property of DPPs is that while they encourage diversity, they do not control for the size of the sampled subsets. In many applications, however, one needs to sample subsets of a fixed size. $k$-DPPs were introduced to address this exact shortcoming. A $k$-DPP is a DPP conditioned on selecting exactly $k$ elements~\cite{kulesza2011k}. It defines a distribution over all the subsets $Y \subseteq Z$ of size $k$, i.e., $|Y| = k$, like so:
\begin{equation*}
    \mathcal{P}(Y \mid |Y| = k) \propto \det(\mathbf{K}_Y)~\text{.}
\end{equation*}

Sampling from DPPs or $k$-DPPs can be computationally expensive, especially when item sets are large, which is the case in this work. Exact algorithms require eigendecompositions of the kernel matrix, a process that has $\mathcal{O}(n^3)$ complexity. To overcome this challenge, several approximate or efficient methods have been proposed, including low-rank approximations and greedy MAP inference~\cite{kulesza2012determinantal, gartrell2017low}.
Further technical details, proofs and more advanced formulations of DPPs and $k$-DPPs can be found in the literature~\cite{kulesza2012determinantal,kulesza2011k}.

\section{Theoretical Complexity of DivNS}%
\label{apx:tc}

In Section~\ref{sec:dpp}, we briefly discussed the theoretical complexity of our diversity-augmented sampler. Here, we provide a more detailed calculation. %
Specifically, there are three factors contributing to the complexity of our bespoke $k$-DPP. %
\begin{enumerate}[leftmargin=*,topsep=0pt]
    \item 
We need to compute the penalty factor $q_i$ -- given by Equation~\ref{eq:penalty_factor} -- for each candidate item $i \in C_u$. %
Since the item embeddings are pre-normalised, this reduces to a matrix product between the $d$-dimensional embeddings of items in $C_u$ and $H_u$. This step therefore costs $\mathcal{O}(|C_u| \times |H_u| \times d)$. %
    \item
Constructing the kernel matrix $\mathbf{K}_{C_u}$ -- from which the sub-matrices $\mathbf{K}_D$ needed in Equation~\ref{eq:dpp-prob} are extracted -- costs $\mathcal{O}({|C_u|}^2 \times d)$. %
Scaling each entry of this matrix with $q_i q_j$ as given by the augmented kernel function $\widetilde{\kappa}$
requires $\mathcal{O}({|C_u|}^2)$ operations. The total cost of constructing the final kernel matrix is thus $\mathcal{O}({|C_u|}^2 \times d + {|C_u|}^2) = \mathcal{O}({|C_u|}^2 \times (d + 1)) = \mathcal{O}({|C_u|}^2\times d )$. 
    \item
Sampling a subset of size $k=|D_u|=|I_u^{+}|$ from $C_u$ using a greedy MAP-based approximation of $k$-DPP %
costs $\mathcal{O}(|I_u^{+}| \times {|C_u|}^2)$. %
\end{enumerate}

The overall complexity of our sampler is thus $\mathcal{O}(|C_u|\times |H_u| \times d + {|C_u|}^2\times d + |I_u^{+}|\times {|C_u|}^2)$.  %
Since $|H_u| = |I^+_u|$, this expression becomes $\mathcal{O}(|C_u|\times |I_u^+| \times d + {|C_u|}^2\times d + |I_u^{+}|\times {|C_u|}^2) = \mathcal{O}({|C_u|}^2 \times (|I_u^{+}| + d)) + \mathcal{O}(|C_u|\times |I_u^+| \times d )$;
and because $|C_u|$ dominates $|I_u^{+}|$, it further simplifies to $\mathcal{O}({|C_u|}^2 \times (|I_u^{+}| + d)$. %
In general, $|I_u^{+}|$ is small because recommendation datasets are sparse and $d$ is the fixed item embedding size. %
Given that $|C_u|=m \times |I_u^{+}|$,
keeping the cache ratio $m$ small speeds up our technique.

\begin{table}[b]
    \centering 
    \footnotesize
    \setlength{\tabcolsep}{2.55pt}%
    \caption{Dataset statistics.}\label{tab:dataset}
    \begin{tabular}{@{}lrrrrl@{}}  
        \toprule
        Dataset & $|\mathcal{U}|$ & $|\mathcal{I}|$ & $|\mathcal{Y}|$ & Density & Characteristics\\ \midrule
        Amazon Beauty~\cite{he2016ups} & 22,363 & 12,101 & 198,502 & 0.07\% & small \& sparse  \\
        ML 1M~\cite{harper2015movielens} & 6,040 & 3,706 & 1,000,209  & 4.47\% & medium \& dense \\
        Pinterest~\cite{geng2015learning} & 55,187 & 9,916 & 1,463,581 & 0.27\% & large \& dense  \\
        Yelp 2022~\cite{yelp_open_dataset}   & 287,116  & 148,523 & 4,392,169 &  0.01\% & large \& sparse \\
        \bottomrule
        \end{tabular}
\end{table}

\section{Experiments}%

\subsection{Setup}\label{app:setup}%

\begin{figure}[b]
    \centering
    \begin{subfigure}{0.48\linewidth}
        \includegraphics[width=\linewidth]{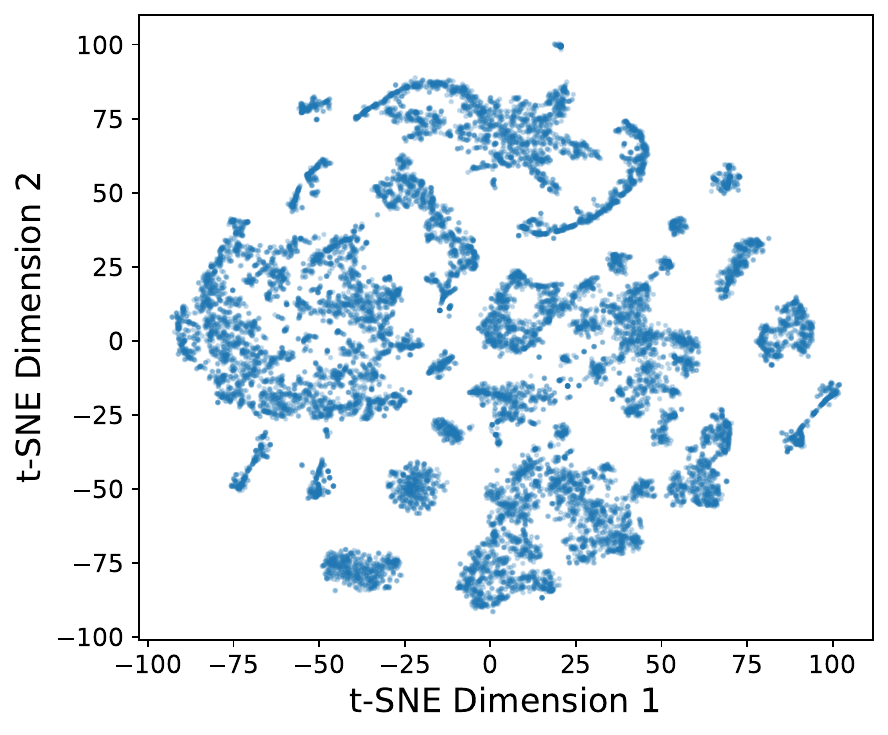}
        \caption{Pinterest.}
    \label{fig:dataset-tsne:p}
    \end{subfigure}
    \hfill
    \begin{subfigure}{0.48\linewidth}
        \includegraphics[width=\linewidth]{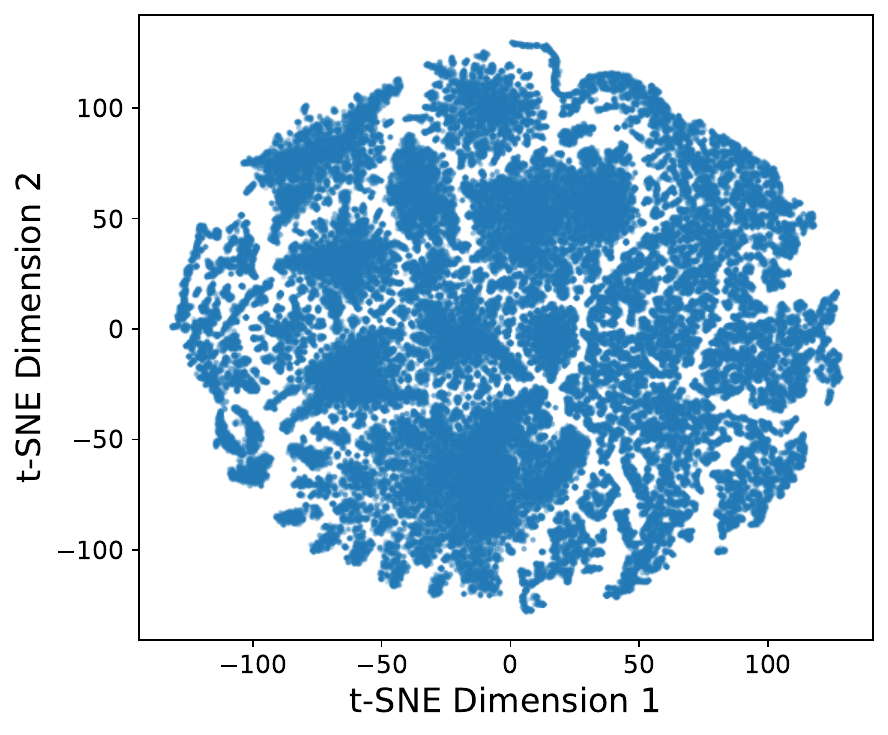}
        \caption{Yelp 2022.}
    \label{fig:dataset-tsne:y}
    \end{subfigure}
    \caption{Visualisation of item embeddings for two benchmark datasets -- (\subref{fig:dataset-tsne:p})~Pinterest and (\subref{fig:dataset-tsne:y})~Yelp 2022 -- using t-SNE. A clear cluster structure can be observed in both cases.}
    \label{fig:dataset-tsne}
\end{figure}

\paragraph{Datasets}
Table~\ref{tab:dataset} shows the statistics of the datasets used in our experiments. Data density is computed as $\frac{|\mathcal{Y}|}{|\mathcal{U}|\times|\mathcal{I}|}$; dataset characteristics are based on prior work~\cite{chin2022datasets}. %
Our method offers most benefit when dataset items %
exhibit a natural cluster structure -- e.g., due to item popularity or genre -- which holds for all the datasets listed in Table~\ref{tab:dataset}. %
To demonstrate this phenomenon, %
Figure~\ref{fig:dataset-tsne} uses t-SNE to visualise learnt item embeddings for Pinterest and Yelp 2022, %
which both exhibit a clear cluster structure. %

\paragraph{Training}
We used early stopping based on Recall@20, terminating training if the metric stopped improving for 20 consecutive epochs. %
We capped the epoch number at 1,000. %
We used LightGCN with three layers.  %

\paragraph{Hyperparameters}
Hyperparameters were optimised with grid search. %
The learning rate $lr$ was searched in $\{0.0001,0.0005,0.001\}$ and the $l_2$ regularisation weight was tuned over $\{10^{-5}, 10^{-4}, 10^{-3}\}$. %
The candidate size $n$ used with DNS, DNS(M, N), MixGCF, DENS and DivNS was searched in $\{6,8,10,15,20\}$. 
The cache ratio $m$ of DivNS was tuned  over $\{1,2,4,6\}$. %
The mixing coefficient $\lambda$ was searched in $\{0.1, 0.3, 0.5, 0.7, 0.9\}$. 
Table~\ref{tab:app-hyper} provides the final configuration of DivNS across all the experiments. %

\begin{table}[t]
    \centering
    \footnotesize
    \setlength{\tabcolsep}{2.55pt}%
    \caption{Hyperparameter setup of DivNS.}
    \begin{tabular}{@{}llrrrrr@{}}
    \toprule
         & & $lr$ & $l_2$ & $n$ & $m$ & $\lambda$ \\\midrule
    \multirow{4}{*}{\rotatebox[origin=c]{90}{\parbox[c]{1cm}{\centering MF}}}  
       & Amazon Beauty & $10^{-3}$ & $10^{-4}$ & 10 & 4 & 0.7 \\
       & ML 1M         & $10^{-3}$ & $10^{-3}$ & 10 & 6 & 0.7 \\
       & Pinterest     & $10^{-4}$ & $10^{-3}$ & 10 & 4 & 0.5 \\
       & Yelp 2022     & $10^{-3}$ & $10^{-3}$ & 10 & 2 & 0.7
       \\ \midrule
    \multirow{4}{*}{\rotatebox[origin=c]{90}{\parbox[c]{1cm}{\centering LightGCN}}} 
       & Amazon Beauty & $10^{-4}$ & $10^{-4}$ & 10 & 4 & 0.7 \\
       & ML 1M         & $10^{-3}$ & $10^{-3}$ & 10 & 4 & 0.7 \\
       & Pinterest     & $10^{-3}$ & $10^{-3}$ & 10 & 4 & 0.5 \\
       & Yelp 2022     & $10^{-3}$ & $10^{-3}$ & 10 & 4 & 0.5 \\
    \bottomrule
    \end{tabular}
    \label{tab:app-hyper}
\end{table}

\subsection{Results}\label{app:add-results}%

\begin{table}[b]
    \centering
    \footnotesize
    \setlength{\tabcolsep}{2.55pt}%
    \caption{Diversity of negatives sampled by each NS technique during LightGCN training. The best results are in bold.}%
    \begin{tabular}{@{}lrrrr@{}}
    \toprule
    & Amazon Beauty & ML 1M & Pinterest & Yelp 2022 \\\midrule
    RNS & 0.136 & 0.156 & 0.158 & 0.183\\
    PNS & 0.092 & 0.149 & 0.142 & 0.142\\
    DNS & 0.153 & 0.170 & 0.156 & 0.127  \\
    AdaSIR & 0.145 & 0.134 & 0.103 & 0.164\\
    DNS(M,N) & 0.138 & 0.178 & 0.095 & 0.195 \\
    MixGCF & 0.114 & 0.192 & 0.133 & 0.185 \\
    DENS & 0.129 & 0.157 & 0.113 & 0.175 \\
    AHNS & 0.104 & 0.147 & 0.144 & 0.183 \\
    SCONE & 0.107 & 0.110 & 0.183 & 0.119 \\
    DivNS & \textbf{0.230} & \textbf{0.248} & \textbf{0.222} & \textbf{0.246} \\
    \bottomrule
    \end{tabular}
    \label{tab:app-diversity}
\end{table}

\paragraph{Runtime}%
The experiments were executed on an NVIDIA P100 GPU with 16GB of VRAM. %
The average training times per epoch were:
32, 158, 222, 284 and 282 seconds respectively for NS, DNS, MixGCF, DENS and DivNS on ML 1M; %
35, 176, 248, 272 and 280 seconds on Pinterest; and %
50, 218, 306, 376 and 413 seconds on Yelp. %
Our results show that even though DivNS requires slightly more time due to its $k$-DPP sampling, the training time remains competitive when compared to other negative samplers. %

\paragraph{Sampled Negatives Diversity}%
Table~\ref{tab:app-diversity} shows the complete results of sample diversity for different NS methods used with LightGCN across all the datasets; %
it complements the result reported in Section~\ref{sec:result} and Figure~\ref{fig:different-div}. %
We observe that DivNS consistently achieves best diversity across the board. %

\begin{figure}[b]
    \centering
    \begin{subfigure}{0.48\linewidth}
        \includegraphics[width=\linewidth]{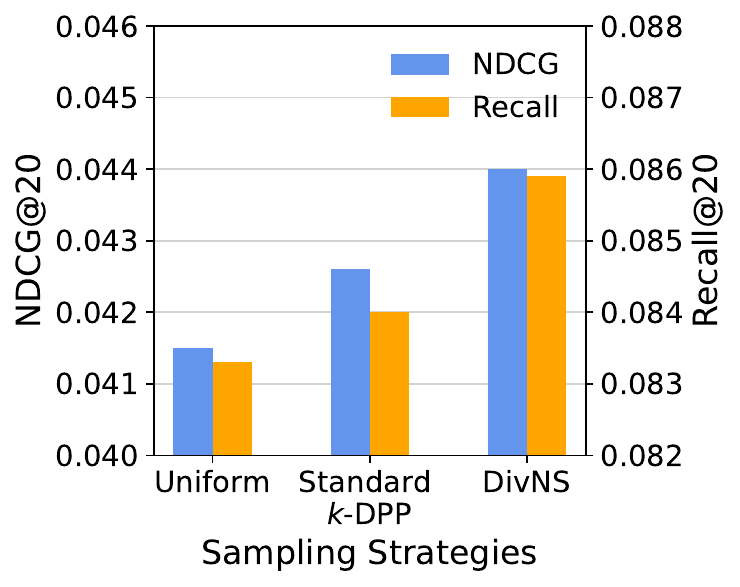}
        \caption{Amazon Beauty.}
        \label{fig:beauty-sampling}
    \end{subfigure}
    \hfill
    \begin{subfigure}{0.48\linewidth}
        \includegraphics[width=\linewidth]{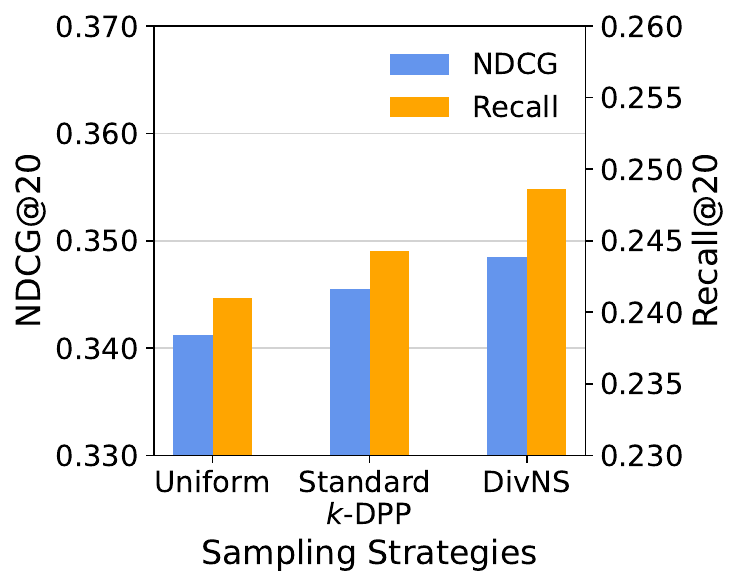}
        \caption{ML 1M.}
        \label{fig:ml1m-sampling}
    \end{subfigure}
    \\
    \begin{subfigure}{0.48\linewidth}
        \includegraphics[width=\linewidth]{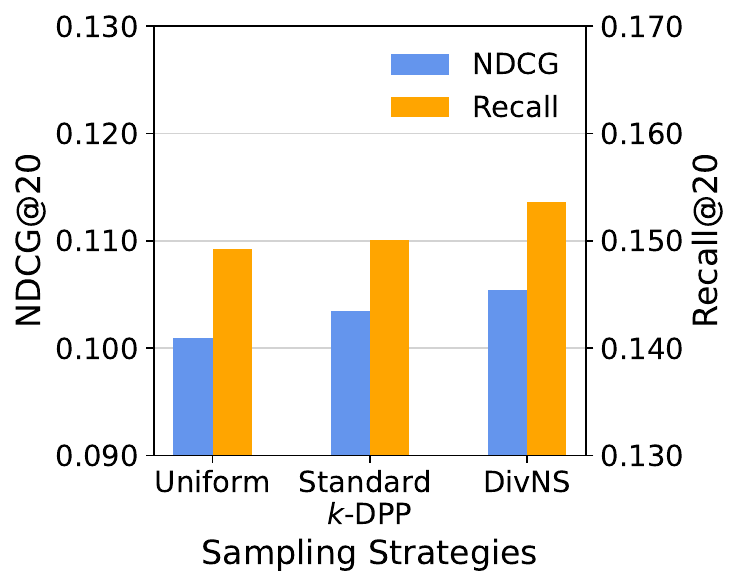}
        \caption{Pinterest.}
        \label{fig:pinterest-sampling}
    \end{subfigure}
    \hfill
    \begin{subfigure}{0.48\linewidth}              
        \includegraphics[width=\linewidth]{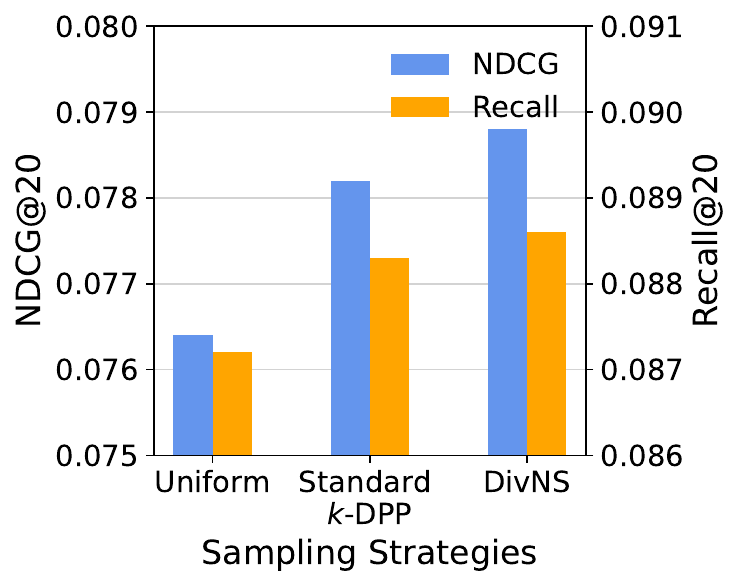}
        \caption{Yelp 2022.}
        \label{fig:yelp-sampling}
    \end{subfigure}
    \caption{Impact of diversity-augmented sampling on performance -- NDCG@20 and Recall@20 -- of LightGCN for: (\subref{fig:beauty-sampling})~Amazon Beauty, (\subref{fig:ml1m-sampling})~ML 1M, (\subref{fig:pinterest-sampling})~Pinterest and (\subref{fig:yelp-sampling})~Yelp 2022.}%
    \label{fig:ablation-sampling-complete}
\end{figure}

\paragraph{Diversity-Augmented Sampling}%
Figure~\ref{fig:ablation-sampling-complete} reports additional results for the experiments evaluating impact of diversity-augmented sampling given in Section~\ref{sec:ablation}. %
It shows that %
our approach consistently outperforms other sampling strategies. %

\begin{figure}[b]
    \centering
    \begin{subfigure}{0.48\linewidth}
        \includegraphics[width=\linewidth]{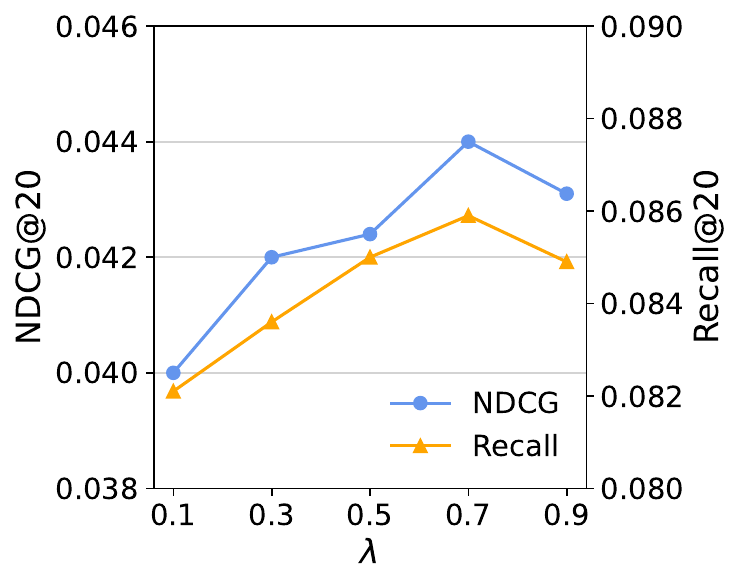}
        \caption{Amazon Beauty.}
        \label{fig:beauty-lambda}
    \end{subfigure}
    \hfill
    \begin{subfigure}{0.48\linewidth}
        \includegraphics[width=\linewidth]{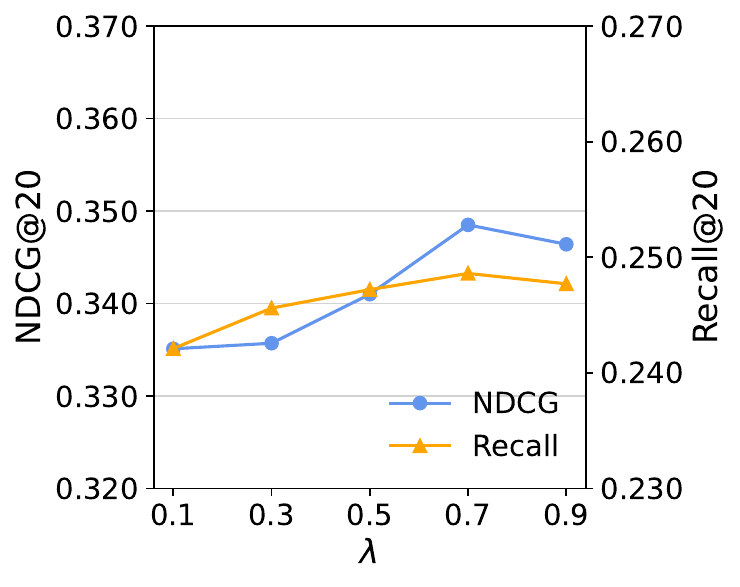}
        \caption{ML 1M.}
        \label{fig:ml1m-lambda}
    \end{subfigure}
    \caption{Impact of synthetic negatives generation on performance -- NDCG@20 and Recall@20 -- of LightGCN for: (\subref{fig:beauty-lambda})~Amazon Beauty and (\subref{fig:ml1m-lambda})~ML 1M.}%
    \label{fig:ablation-lambda-complete}
\end{figure}

\paragraph{Synthetic Negatives Generation}%
Figure~\ref{fig:ablation-lambda-complete} reports results on two additional datasets for the experiments evaluating the impact of synthetic negatives generation given in Section~\ref{sec:ablation}. %
These findings align with our original conclusion that diverse negatives can help model generalisation, thus improve performance, but, at the same time, their excessive amount can be detrimental. %

\begin{table}[b]
    \centering
    \footnotesize
    \setlength{\tabcolsep}{2.55pt}%
    \caption{Impact of cache ratio $m$ on DivNS performance measured with NDCG@20 and Recall@20. We fix $n=10$. The best results are in bold.
    }
    \begin{tabular}{@{}lrrrrrrrrr@{}}
    \toprule
     & & \multicolumn{2}{c}{Amazon Beauty} & \multicolumn{2}{c}{ML 1M} & \multicolumn{2}{c}{Pinterest} & \multicolumn{2}{c}{Yelp 2022}  \\
     \cmidrule(lr){3-4} \cmidrule(lr){5-6} \cmidrule(lr){7-8}
     \cmidrule(lr){9-10}  
     & & NDCG & Recall & NDCG & Recall & NDCG & Recall & NDCG & Recall \\ \midrule
     \multirow{4}{*}{\rotatebox[origin=c]{90}{\parbox[c]{1cm}{\centering MF}}} & $m$=1 & 0.0359 & 0.0705 & 0.3320 & 0.2284 & 0.0830 & 0.1285 & 0.0602 & 0.0713  \\
     & $m$=2 & 0.0362 & 0.0743 & 0.3325 & 0.2341 & 0.0852 & 0.1316 & 0.0613 & \textbf{0.0727} \\
     & $m$=4 & \textbf{0.0385} & \textbf{0.0758} & 0.3353 & \textbf{0.2408} & \textbf{0.0883} & \textbf{0.1356} & \textbf{0.0615} & 0.0722 \\
     & $m$=6 & 0.0378 & 0.0733 & \textbf{0.3397} & 0.2393 & 0.0861 & 0.1362 & 0.0609 & 0.0714 \\
     \midrule
     \multirow{4}{*}{\rotatebox[origin=c]{90}{\parbox[c]{1cm}{\centering LightGCN}}} & $m$=1 & 0.0431 & 0.0844 & 0.3467 & 0.2445 & 0.0990 & 0.1464 & 0.0762 & 0.0870 \\
     & $m$=2 & 0.0437 & 0.0845 & 0.3471 & 0.2460 & 0.0994 & 0.1480 & 0.0774 & 0.0872 \\
     & $m$=4 & \textbf{0.0440} & \textbf{0.0859} & 0.3485 & \textbf{0.2486} & \textbf{0.1054} & \textbf{0.1536} & \textbf{0.0788} & \textbf{0.0886} \\
     & $m$=6 & 0.0429 & 0.0850 & \textbf{0.3502} & 0.2475 & 0.1047 & 0.1515 & 0.0783 & 0.0878 \\
    \bottomrule
    \end{tabular}
    \label{tab:ablation-cache-size}
\end{table}

\paragraph{Cache Ratio}%
Table~\ref{tab:ablation-cache-size} shows the performance of DivNS for different cache ratio setups with the candidate set size set as $n=10$.

\end{document}